# *FieldSeer I: Physics-Guided World Models for Long-Horizon Electromagnetic Dynamics under Partial Observability*

Ziheng Guo[1*], Fang Wu[2], Maoxiong Zhao[2], Chaoqun Fang[2], Yang Bu[1*]
[1, *]Zhangjiang Laboratory, Shanghai, China

## ABSTRACT

We introduce FieldSeer I, a geometry-aware world model that forecasts electromagnetic field dynamics from partial observations in 2-D TE waveguides. The model assimilates a short prefix of observed fields, conditions on a scalar source action and structure/material map, and generates closed-loop rollouts in the physical domain. Training in a symmetric-log domain ensures numerical stability. Evaluated on a reproducible FDTD benchmark (200 unique simulations, structure-wise split), FieldSeer I achieves higher suffix fidelity than GRU and deterministic baselines across three practical settings: (i) software-in-the-loop filtering (64×64, P=80→Q=80), (ii) offline single-file rollouts (80×140, P=240→Q=40), and (iii) offline multi-structure rollouts (80×140, P=180→Q=100). Crucially, it enables edit-after-prefix geometry modifications without re-assimilation. Results demonstrate that geometry-conditioned world models provide a practical path toward interactive digital twins for photonic design.

## Acknowledgment

This paper presents FieldSeer I, a minimal prototype for geometry-aware EM forecasting in 2D waveguides. While focused on reproducible benchmarking, an enhanced version with 3D support and multi-physics capabilities is already deployed in production engineering workflows. This research was supported by computational resources provided by Zhangjiang Laboratory under project No. 202320202-2. The technology described in this paper has been submitted for patent protection (Application No.: CN202511746268.4). Due to pending patents and proprietary adaptations, we disclose only the foundational architecture and evaluation protocol described herein. Full implementation details of advanced variants may be subject to proprietary constraints and will be disclosed following institutional policies and patent approval procedures.

## 1 INTRODUCTION

The integration of machine learning into nanophotonics has enabled fast surrogate models that replace costly full-wave simulations by learning mappings from geometry and excitation conditions to electromagnetic responses [1–4,7,10]. Yet most of these approaches are static: they predict a spectrum or steady-state field for a fixed structure, offering no mechanism to forecast temporal dynamics or respond to mid-sequence interventions. This limits their utility in interactive design settings where one observes a short field history, modifies the geometry, and wishes to immediately preview future behavior—without restarting a solver.

Recent advances in learning-based PDE solvers—such as neural operators [5–6] and physics-informed networks [8–9]—improve data efficiency or physical consistency but still struggle with long-horizon stability, partial observability, and rapid adaptation to structural changes. Meanwhile, world models from reinforcement learning maintain latent states and "imagine" future trajectories under actions [11–14], yet their application to EM dynamics with explicit geometry conditioning remains largely unexplored.

We present FieldSeer I, a geometry-aware, action-conditioned world model for 2-D TE waveguide dynamics. Given a short prefix of observed fields, it conditions on both a scalar source action and a full structure/material map, and generates closed-loop rollouts of future fields. Critically, the model supports on-the-fly geometry edits after the prefix: modifications to the structure take immediate effect in subsequent predictions without re-assimilation or solver restarts. To ensure numerical stability across varying field amplitudes and long horizons, training is performed in a symmetric-log (symlog) domain, while all metrics are reported in the physical domain after inverse transformation. We also explore a lightweight online adaptation step—updating the model on a single held-out sequence—to mimic "software-in-the-loop" usage without full retraining.

We evaluate FieldSeer I on a reproducible 2-D FDTD benchmark with structure-wise train/validation/test splits, ensuring test geometries are unseen. Using a prefix→suffix protocol (e.g., 64×64: P=80, Q=80), we report physical-domain PSNR and MSE on suffix frames. Baselines include GRU and prototype architectures trained under identical data, resolution, and early-stopping conditions. Results demonstrate that FieldSeer I enables editable, dynamics-aware forecasting in partially observed photonic environments—a step toward interactive, simulation-free design.

## 2. FieldSeer I

### 2.1 Data Generation

We generate our dataset by simulating 2D transients on a Yee grid using leap-frog updates under a CFL-stable timestep. In vacuum cells, the semi-discrete curl equations govern field evolution:

$$\frac{\partial E}{\partial t} = \nabla \times H, \frac{\partial H}{\partial t} = -\nabla \times E \qquad (1)$$

Dispersive pixels follow a single-pole Lorentz dielectric model:

$$\varepsilon(\omega) = \varepsilon_\infty + \frac{f\omega_0^2}{\omega_0^2 - \omega^2 - i\omega\gamma} \qquad (2)$$

with parameters fitted from $n$, $k$ tables within a narrow band around the target wavelength. Excitation is introduced via a band-limited Gaussian pulse injected through a total-field/scattered-field (TFSF) interface, ensuring scattered fields from embedded inclusions propagate cleanly into the observation region. To suppress wrap-around reflections, we combine a thin perfectly matched layer (PML) with Mur-type boundary updates. Along the propagation axis (x), we place two 1D probes — one upstream ("reflection"), one downstream ("transmission") — and log both $E_z$ and $H_y$. ower is computed via the Poynting component $S_x = E_z H_y$ (TE + x propagation). Geometries consist of multiple non-overlapping circular inclusions placed between the probes and away from PML regions, with radii and counts sampled per episode. For each grid resolution (e.g., 64×64, 76×140, 128×256, etc; where 64 units correspond to 0.64 μm), the physical pixel size Δx is derived from the domain size; PML thickness scales proportionally to maintain absorption efficiency. TFSF slab indices and probe positions are stored as metadata for reproducibility. A full transient of length $T$ is simulated at a fixed λ (e.g., 1550 nm), with the instantaneous $E_z$ field recorded at every timestep.

## 2.2 Model Architecture

We consider the problem of long-horizon forecasting of a 2D electromagnetic field $E_z^t \in \mathbb{R}^{H \times W}$ driven by a scalar source amplitude $a^t$. A static structure/material map $STR \in \mathbb{R}^{C_{str} \times H \times W}$ (e.g., inclusions, radii, material tags, $n$, $k$, etc) is provided for the entire rollout. The model has five components: (i) a stable field representation, (ii) geometry-aware tokenization and Transformer plan context, (iii) recurrent dynamics with a structural drive (GRU core), (iv) a lightweight stochastic latent for uncertainty and generalization, and (v) a structure-conditioned decoder enabling editable rollouts. Figure 1 shows the model architecture.

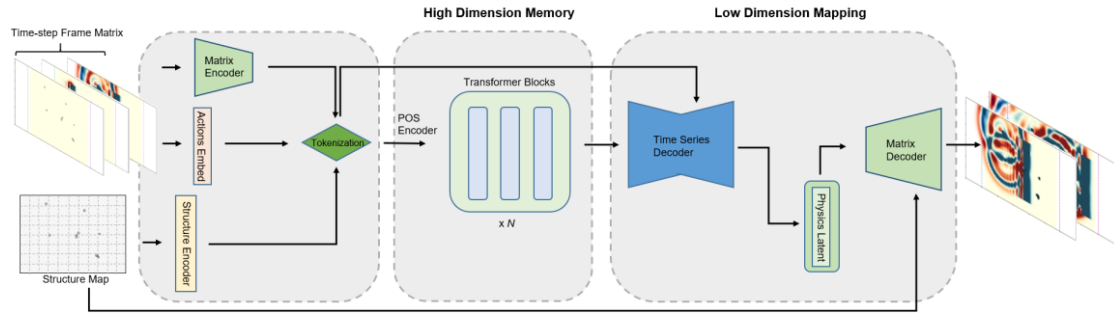

Figure 1. Architecture overview. The model processes a time-step frame matrix and a structure map. The frame matrix is encoded by a Matrix Encoder, while time-step actions (e.g., light source timing) are embedded via an Actions Embedding module. The structure map is encoded separately. All features are concatenated, tokenized, and enriched with learnable positional encodings before being processed by N Transformer Blocks. The high-dimensional output is decoded by a Time Series Decoder, projected into a Physics Latent space, and reconstructed via a Matrix Decoder. The final prediction is fused with the original structure map to preserve spatial constraints and generate the predicted field.

### 2.2.1 Stable Field Representation via Symlog Transform

Electromagnetic fields can span several orders of magnitude. We therefore work in a symmetric log domain:

$$\tilde{E}^t = \text{symlog}(E_z^t) = \text{sign}(E_z^t) \cdot \log(1 + |E_z^t|) \tag{3}$$

with inverse $symexp(x) = sign(x)(\exp(|x|) - 1)$ for visualization. We do not apply z-score normalization; empirical tests show similar performance, but symlog is chosen for its ability to stabilize gradient propagation during long-horizon training by compressing the dynamic range of field amplitudes. Additional stability arises from per-step normalization of the recurrent state.

### 2.2.2 Geometry-Aware Tokenization and Transformer Context

We encode fields and structure into a common $D$ dimensional space:

$$e^t = f_\phi(\tilde{E}^t) \in \mathbb{R}^D, \qquad u^t = W_a a^t \in \mathbb{R}^D, \qquad s = f_{str}(STR) \in \mathbb{R}^D \tag{4}$$

Structure/material is injected at the token level by shifting both observation and action tokens:

$$o_t = e^t + W_{se}s, \qquad \tilde{a}_t = u^t + W_{sa}s \tag{5}$$

We construct an alternating, causal token stream of length $2U$:

$$x_{2t} = o_t, \qquad x_{2t-1} = \tilde{a}_t, \qquad t = 0, \dots, U-1 \tag{6}$$

apply learnable positional encodings, and pass through a Transformer with a causal mask $M_{ij} = \mathbf{1}[j \leq i]$:

$$h_k^0 = x_k + p_k, \qquad h^L = T(h^0; M) \tag{7}$$

We extract a plan context at each step from the action positions:

$$c_t = h_{2t-1}^L \in \mathbb{R}^D \tag{8}$$

### 2.2.3 Recurrent Dynamics with Structural Drive

A GRU consumes the plan context $c_t$, the previous latent $z_{t-1}$, optional pixel feedback $e^{t-1}$, and a structural drive $W_s s$:

$$r^t = c_t + W_z z^{t-1} + \underbrace{W_e e^{t-1}}_{optional} + W_s s,$$
$$\tilde{h}^t = \text{GRU}(h_{t-1}, r^t), h_t = Norm(\tilde{h}^t) \quad (9)$$

Pixel feedback uses ground-truth features during the prefix and model predictions thereafter:

$$e^t = \begin{cases} f_\phi(\tilde{E}^t), t < P \\ f_\phi\left(symlog(\hat{E}^t)\right), t > P \end{cases} \quad (10)$$

### 2.2.4 Stochastic Latent for Uncertainty and Generalization

We use a diagonal-Gaussian latent $z_t \in \mathbb{R}^Z$ with posterior on the prefix and prior thereafter:

$$q_\theta(z_t | h_t, \tilde{E}^t) = \mathcal{N}(\mu_q\left([h_t; vec(\tilde{E}^t)], \sigma_q(\cdot)\right) \quad (11)$$

$$p_\theta((z_t | h_t) = \mathcal{N}\left(\mu_q(h_t), \sigma_q(h_t)\right) \quad (12)$$

Usage. During training one may sample from $q$ (or take its mean) on the prefix; at inference we typically take the mean of $q$ for $t<P$ and the mean of $p$ for $t\geq P$, then feed to the decoder.

### 2.2.5 Structure-Conditioned Decoder for Editable Rollouts

The reconstruction uses both the dynamical state and the raw structure map as conditioning channels:

$$\hat{E}^t = d_\emptyset([h_t; z_t], STR) \in \mathbb{R}^{H \times W} \quad (13)$$

Because structure enters tokens, dynamics, and the decoder, edits to $STR$ after the prefix immediately affect future rollouts—no re-assimilation or solver restart is required.

**Training objective**

Prefix reconstruction, open-loop prediction, an rFFT2 auxiliary, and we adopt Dreamer-style balanced KL with free-nats:

$$\mathcal{L}_{\text{rec}} = \frac{1}{P} \sum_0^{t-1} \left\| symlog(\hat{E}^t) - symlog(E^t) \right\|_2^2,$$

$$\mathcal{L}_{\text{pred}} = \frac{1}{U-P} \sum_{t=P}^{U-1} \left\| symlog(\hat{E}^t) - symlog(E^t) \right\|_2^2 \tag{14}$$

$$\mathcal{L}_{\text{spec}} = \frac{1}{U-P} \sum_{t=P}^{U-1} \left( \left\| \mathcal{R}(\mathcal{F}\hat{E}^t) - \mathcal{R}(\mathcal{F}E^t) \right\|_2^2 + \left\| \mathfrak{I}(\mathcal{F}\hat{E}^t) - \mathfrak{I}(\mathcal{F}E^t) \right\|_2^2 \right) \tag{15}$$

$$\mathcal{L}_{KL}^t = aKL(q_\theta(z_t|h_t,\tilde{E}^t) \| sg\ p_\theta(z_t|h_t)) + (1-a)KL(sg\ q_\theta(z_t|h_t,\tilde{E}^t) \| p_\theta(z_t|h_t)) \tag{13}$$

$$\mathcal{L}_{\text{KL}} = \frac{1}{P} \sum_{0}^{t-1} \max(\mathcal{L}_{KL}^t, \varepsilon) \tag{16}$$

$$\mathcal{L}_{\text{total}} = \lambda_{rec}\mathcal{L}_{rec} + \lambda_{pred}\mathcal{L}_{pred} + \lambda_{spec}\mathcal{L}_{spec} + \lambda_{KL}\mathcal{L}_{KL} + \underbrace{\lambda_{PIF}\mathcal{L}_{PIF}}_{optionally\ PINN} \tag{17}$$

Optionally Physics-inspired latent/field residuals remain available.

### 3.1 Experimental setup

**Datasets and splits.** We use the FDTD-generated sequences described in Section 2.1. Each example provides a static structure/material map ($STR$), an action stream $a_t$, and a field sequence $E_z^t$ of length $U=P+Q$. We split by structure (train/val/test), ensuring that no structure appears in more than one split. **Dataset scale**. To demonstrate data efficiency, the dataset comprises 200 unique simulations; we use 180 for training and 20 for validation/testing. **Note**. Simulations contains different size geometries.

**Training protocol**. We assimilate a prefix of length $P$ with the posterior $q(z_t | h_t, vec(\tilde{E}^t))$ and imagine $Q$ steps with the prior $p(z_t|h_t)$ under the provided action stream. Unless otherwise noted, decoding conditions on the raw structure map (pixel-level channels). We train in the $symlog$ domain and decode back to the physical domain via $symexp$ for evaluation. The single-sample update fine-tunes only the decoder and latent projection layers using ground-truth suffix fields, with a learning rate of 1e-3 for 500 epochs.

**Normalization and evaluation alignment.** All models are evaluated in the physical domain. FieldSeer I is trained in a symmetric-log ($symlog$) domain but predictions are inverse-mapped by $symexp$ before scoring. The GRU baseline outputs z-scored fields and is denormalized using the dataset mean and standard deviation recorded at training time. Prototype outputs linear fields and requires no inverse transform. After these model-specific inverse transforms, we compute metrics on the same frames and pixels for all methods.

**Inference protocol**. In the Transformer stream, observation tokens for $t \geq P$ are zeros to prevent future leakage; the structure code $s$ remains active at all steps. Because geometry/material information is injected at tokens, dynamics, and decoder, editing $STR$ after the prefix immediately propagates to futures—no re-assimilation or solver restart. We follow this protocol for both quantitative evaluation and qualitative figures.

## 3.2 Quantitative results

**Evaluation settings**. We report image-space fidelity using **MSE** (↓) and **PSNR** (↑, dB) in the physical domain. Each model's outputs are inverse-mapped to the physical domain before scoring (FieldSeer: $symexp$ with dataset-fixed $\alpha$; GRU: z-score denormalization using dataset mean/std; Prototype: identity). PSNR uses a **single dataset-wide MAX** (the 99.9th percentile of $|E|$ on the validation set), held fixed across tables. Unless otherwise stated, **suffix-only** metrics are primary (all-frames shown in parentheses).

**Two evaluation modes.**
Software-in-the-Loop (SITL; online filtering, one-step-ahead).
We emulate a deployment where a fast simulator (or instrument) streams frames. The model receives a prefix of $P$ observed frames $x_{1:P}$, then for each suffix step $t = P, \ldots, P + Q - 1$ it is given the current observed frame $x_t$ and asked to predict the next frame $\hat{x}_{t+1}$. We compute one-step-ahead metrics over the $Q$ suffix steps:

$$\text{MSE}_{\text{SITL}} = \frac{1}{QHW} \sum_{t=P}^{P+Q-1} \|\hat{x}_{t+1} - x_{t+1}\|_2^2, \quad \text{PSNR}_{\text{SITL}} = 10 \log_{10} \frac{\text{MAX}^2}{MSE_{SITL}}$$

SITL (Table 1, 64×64, $P$=80, $Q$=80). This one-step-ahead mode emulates deployment: a fast "software" stream provides $x_t$ and the model predicts $x_{t+1}$. We choose a balanced horizon (80/80 of $T$=160) to measure filtering quality without future leakage while keeping the online budget symmetric.

**Offline, single-file (Table 2, 80×140, $P$=240, $Q$=40)**. Here we test whether the model can leave SITL and still roll out in open loop when given just one sequence. We set a long prefix ($P$=240) so the latent has ample assimilation of drive, multipath build-up and boundary interactions; the short suffix ($Q$=40) isolates short-horizon open-loop drift and verifies the model can "load other data" (non-SITL) and remain stable without step-wise observations.

**Offline, multi-file/structure (Table 3, 80×140, $P$=180, $Q$=100)**. This split stresses long-horizon stability and generalization. Compared with single-file, we (i) increase suffix to $Q$=100 to probe error accumulation over many scattering cycles, and (ii) train/evaluate across multiple structures, so success must persist under geometry variation. The longer suffix and multi-file diversity together test whether FieldSeer's closed-loop imagination remains stable and geometry-aware beyond a single trajectory.

**Baselines under both modes.**

All methods run under the same mode as FieldSeer in each table. For SITL (Table 1), GRU and Prototype also ingest $x_t$ each step and predict $\hat{x}_{t+1}$ (we feed $x_t$ through their encoders—no architectural changes). For Offline (Tables 2–3), all methods run open-loop. This keeps comparisons fair despite different training normalizations (symlog vs. z-score vs. linear), because evaluation is unified in the physical domain with a single **MAX** policy.

| Table 1, 64×64, Software-in-the-Loop (SITL) | | |
|---|---|---|
| Model Method | PSNR ↑ (dB) | MSE ↓ |
| FieldSeer I (Newest) | **48.31** | **0.00076** |
| Baseline 2 (GRU) | 35.63 | 0.01406 |
| Prototype | 19.32 | 0.60123 |

Table 1: One-step-ahead PSNR (↑) and MSE (↓) on 64×64 sequences with total $T$=160 steps (prefix $P$=80, suffix $Q$=80, shown in Figure 3.1). At each suffix step the software provides $x_t$; the model predicts $\hat{x}_{t+1}$. Metrics are computed in the physical domain after inverse transforms, with a single dataset-wide MAX (P99.9 of |E|). Ground-truth Variance across data is 0.01297.

| Table 2, 80×140, Offline (single-file) | | |
|---|---|---|
| Model Method | PSNR ↑ (dB) | MSE ↓ |
| FieldSeer I (Newest) | **39.08** | **0.00635** |
| Baseline 2 (GRU) | 24.01 | 0.02041 |
| Prototype | 15.70 | 1.38369 |

Table 2: P=240→Q=40 is chosen to verify the model's ability to operate offline (outside SITL) with a long assimilation window and a short open-loop horizon that isolates early-suffix drift while keeping conditions identical to Fig. 3.2

| Table 3, 80×140, Offline (multi-file / multi-structure). | | |
|---|---|---|
| Model Method | PSNR ↑ (dB) | MSE ↓ |
| FieldSeer I (Newest) | **36.61** | **0.01122** |
| Baseline 2 (GRU) | 20.88 | 0.41980 |
| Prototype | 13.80 | 2.14308 |

Table 3: $P$=180→$Q$=100 stresses long-horizon stability and cross-structure generalization. A longer suffix ($Q$=100) magnifies accumulation effects, while multi-file training/evaluation requires robustness to geometry variation (Fig. 3.3)

### 3.3 Qualitative analysis: field triplets

We include "field triplet" figures at the end of the manuscript. Each triplet comprises (i)

the **ground-truth** field $E_z$ (physical domain), (ii) the model prediction $\hat{E}_z$ (physical domain), and (iii) an absolute error heatmap $|\hat{E}_z - E_z|$. For consistency across timesteps and methods, we use a fixed color scale for fields (centered at 0) and a separate fixed scale for error maps. Where helpful, we overlay the geometry outline (from $STR$) as a thin contour to indicate inclusion boundaries. We typically show one frame at the end of the prefix ($t=P-1$) and two rollout frames (e.g., $t=P+5$ and $t=P+15$). All qualitative examples are drawn from unseen test structures.

**Interpretation**. Uniformly low error near boundaries and within high-index regions indicates that the structure-conditioned decoder is effectively leveraging $STR$. When visible, faint oscillatory residuals tend to align with multiple-reflection paths; these typically diminish with stronger prefix assimilation or spectral auxiliary weight. Failure cases (e.g., elevated errors at sharp corners or in very lossy inclusions) are included for transparency.

## 4 Conclusions

We presented FieldSeer I, a structure-aware world model for electromagnetic dynamics that combines a Transformer-derived plan context, a GRU dynamics core with structural drive, a lightweight stochastic latent, and a structure-conditioned decoder. Training in the symmetric-log domain enables stable long-horizon rollouts with limited supervision. On a compact 2-D FDTD benchmark with strict structure-wise splits, FieldSeer I delivers superior image-space fidelity (PSNR↑/MSE↓) compared to GRU and deterministic baselines, while uniquely supporting edit-after-prefix geometry modifications and single-sample online adaptation.

These capabilities position FieldSeer I as a foundational component for photonic digital twins—interactive simulation environments where engineers can observe partial field dynamics, modify structures mid-simulation, and immediately preview the consequences without restarting full-wave solvers. Future work will focus on scaling to full-vector 3D fields, incorporating multi-physics effects (thermal, nonlinear), and expanding action spaces to support multi-source control and boundary modulation. Crucially, we aim to integrate FieldSeer with response-aware optimization loops for inverse design and close the reality gap through hardware-in-the-loop adaptation on physical measurements. These advances could transform photonic engineering workflows from static simulation-analysis cycles toward truly interactive, dynamics-aware design of optical components and integrated photonic systems.

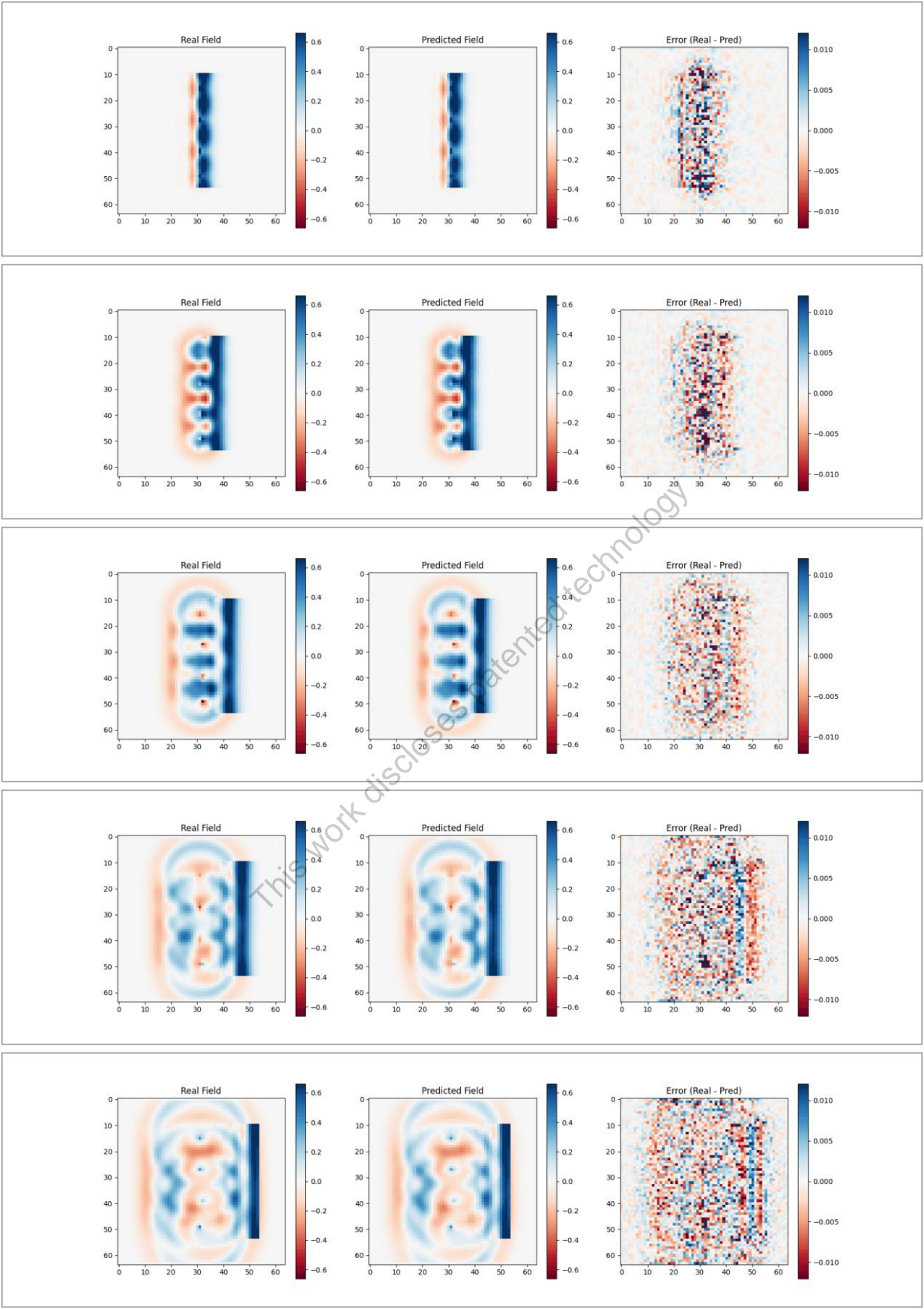

Figure 3.1. 64x64 grid test waveguide data. 80 time-step prefix and 80 time-step predicted field along with error comparison (total 160 time-step).

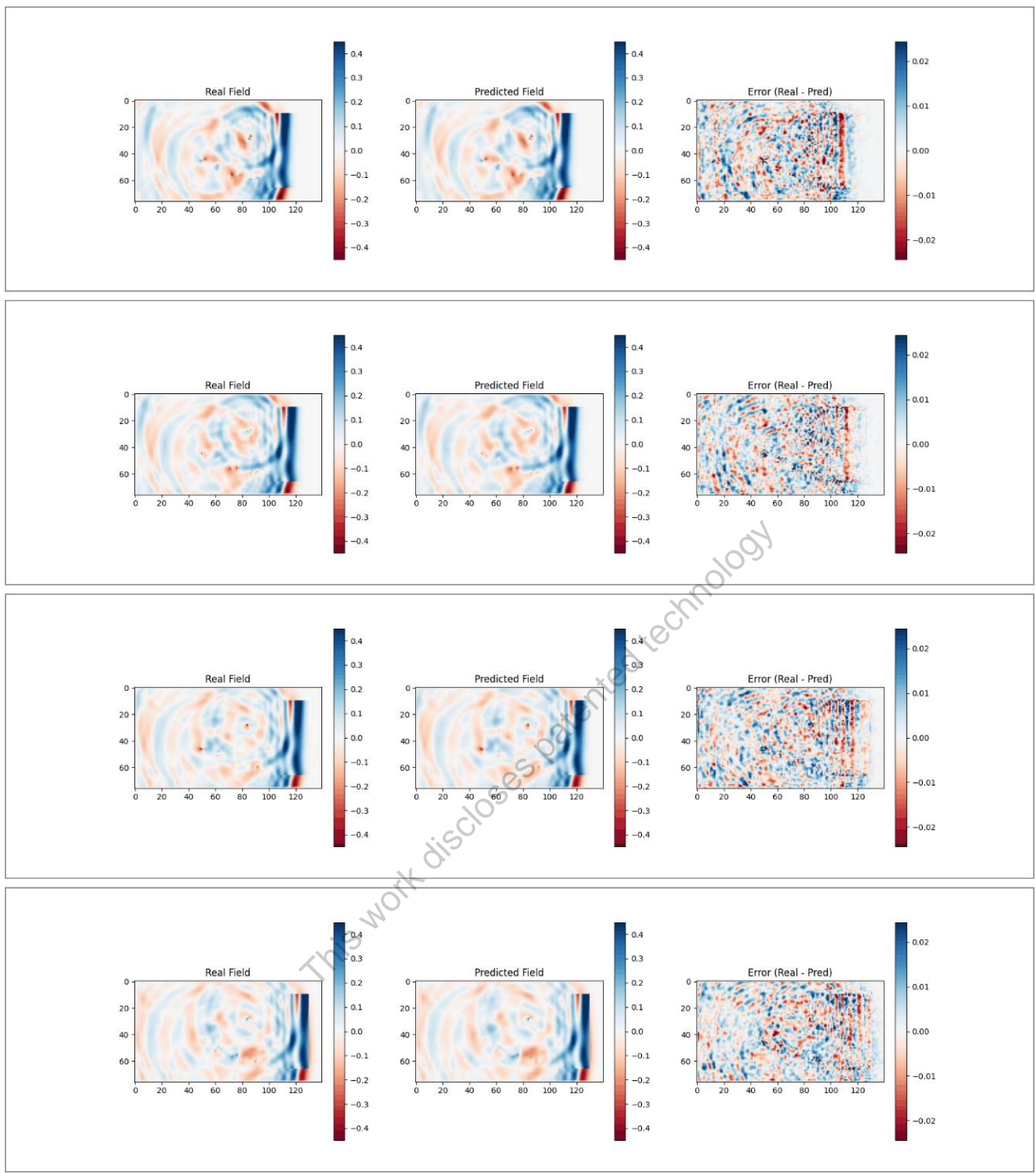

Figure 3.2. Offline rollout after a long prefix (*P*=240) demonstrates that the model can ingest a single sequence and remain stable without per-step observations

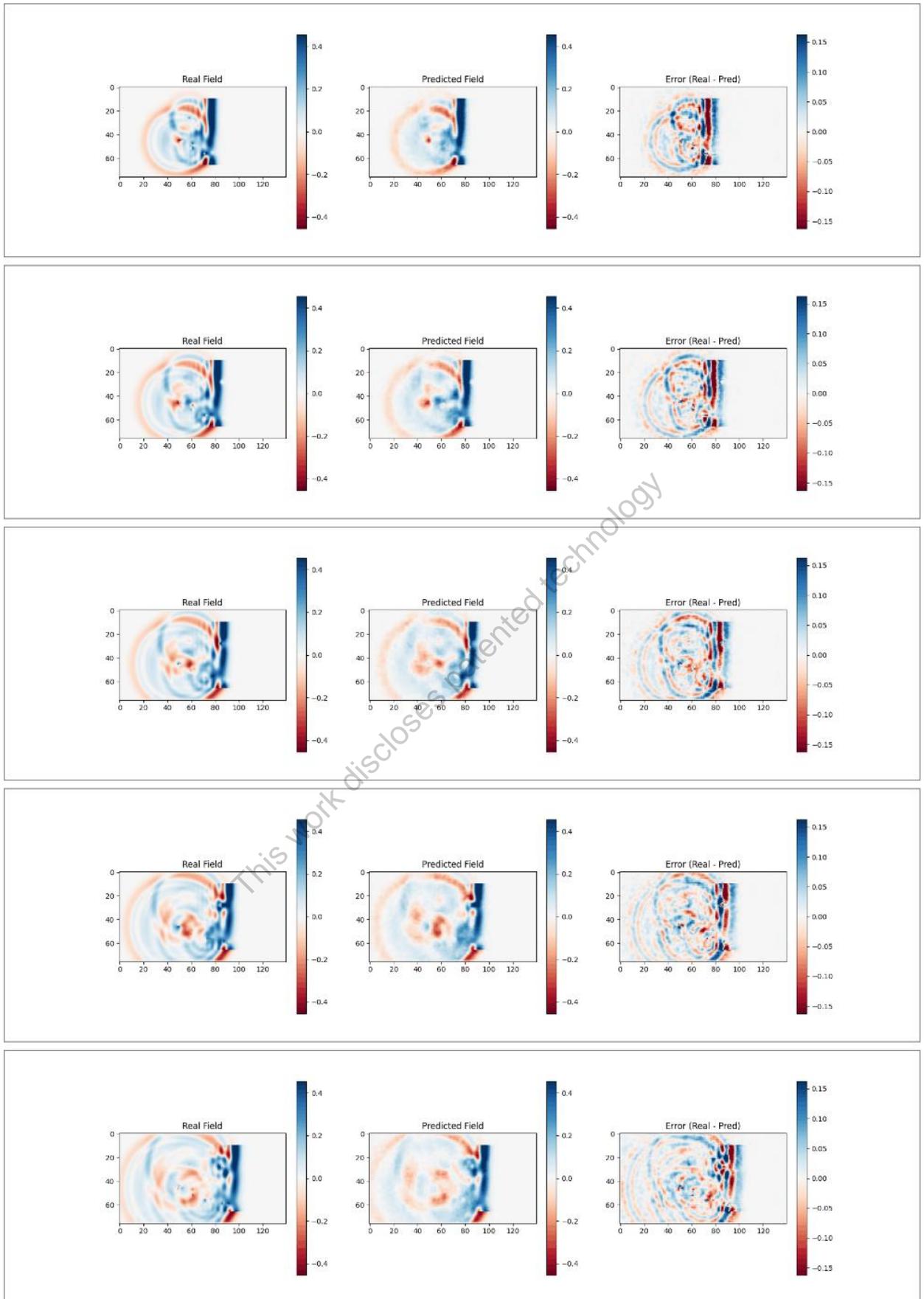

Figure 3.3. Offline rollout with $P$=180] and $Q$=100 over held-out structures highlights long-horizon stability and generalization.

13